\documentclass[11pt]{article}%
\usepackage{amsfonts}
\usepackage{amsmath}
\usepackage{amssymb}
\usepackage{graphicx}
\usepackage{hyperref}%
\providecommand{\U}[1]{\protect\rule{.1in}{.1in}}
\begin{document}

\title{\vspace{-0.5in}Spin Multiplicities}
\author{T. L. Curtright$^{\S }$, T. S. Van Kortryk, and C. K. Zachos$^{\S ,\natural}%
${\small \smallskip}
\and {\small curtright@miami.edu, vankortryk@gmail.com, and
zachos@anl.gov\smallskip}\\$^{\S }$Department of Physics, University of Miami, Coral Gables, FL
33124-8046, USA\\$^{\natural}$High Energy Physics Division, Argonne National Laboratory,
Argonne, IL 60439-4815, USA}
\date{}
\maketitle

\begin{abstract}
The number of times spin $s$ appears in the Kronecker product of $n$ spin $j$
representations is computed, and the large $n$ asymptotic behavior of the
result is obtained. \ Applications are briefly sketched.

\end{abstract}

\section*{Introduction}

We present a derivation of the spin multiplicities that occur in $n$-fold
tensor products of spin-$j$ representations, $j^{\otimes n}$. \ We make use of
group characters, properties of special functions, and asymptotic analysis of
integrals. \ While previous derivations for some of our results are scattered
throughout the literature, especially for specific values of $j$, we provide
here a treatment that is self-contained, and valid for any $j$ and for any
$n$. \ We emphasize two types of novel features: \ patterns that arise when
comparing different values of $j$, and asymptotic behavior for large $n$. \ 

Our methods and results should be useful for various calculations. \ In
particular, the asymptotic behavior that we obtain should be helpful in the
analysis of statistical problems such as the determination of partition
functions. \ In the last section some other applications are briefly
discussed, including a problem of interest for quantum computing, namely, an
estimation of the number of entangled states.

\section*{Basic Theory of Group Characters}

The \href{https://en.wikipedia.org/wiki/Character_theory}{character
$\chi\left(  R\right)  $ of a group representation $R$} succinctly encodes
considerable information about $R$, as is well-known \cite{Texts}. \ For
irreducible representations the characters are orthogonal,%
\begin{equation}
\sum\hspace{-0.2in}\int~\mu~\chi^{\ast}\left(  R_{1}\right)  \chi\left(
R_{2}\right)  =\delta_{R_{1},R_{2}}\ , \label{OrthoChar}%
\end{equation}
where the sum or integral is over the group parameter space with an
appropriate measure $\mu$.

For a Kronecker product of $n$ representations, the character is given by the
product of the individual characters,%
\begin{equation}
\chi\left(  R_{1}\otimes R_{2}\otimes\cdots\otimes R_{n}\right)  =\chi\left(
R_{1}\right)  \chi\left(  R_{2}\right)  \cdots\chi\left(  R_{n}\right)  \ ,
\end{equation}
from which follows an explicit expression for the number of times that a given
representation $R$ appears in the product (e.g., see \cite{V} Chapter I
\S 4.7). \ This multiplicity is
\begin{equation}
M\left(  R;R_{1},\cdots,R_{n}\right)  =\sum\hspace{-0.2in}\int~\mu~\chi^{\ast
}\left(  R\right)  \chi\left(  R_{1}\right)  \chi\left(  R_{2}\right)
\cdots\chi\left(  R_{n}\right)  \ .
\end{equation}
For real characters, this is totally symmetric in $\left\{  R,R_{1}%
,\cdots,R_{n}\right\}  $, and it immediately shows that the number of times
$R$ appears in the product $R_{1}\otimes\cdots\otimes R_{n}$ is equal to the
number of times the trivial or \textquotedblleft singlet\textquotedblright%
\ representation appears in the product $R\otimes R_{1}\otimes\cdots\otimes
R_{n}$.

\section*{The $SU\left(  2\right)  $ Case}

Consider now the Lie group $SU\left(  2\right)  $. \ In this case the
irreducible representations are labeled by angular momentum or spin, $j$ or
$s$, the classes of the group are specified by the angle of rotation about an
axis, $\theta$, and the characters are
\href{https://en.wikipedia.org/wiki/Chebyshev_polynomials#Trigonometric_definition}{Chebyshev
polynomials of the second kind}, $\chi_{j}\left(  \theta\right)
=U_{2j}\left(  \cos\left(  \theta/2\right)  \right)  $. \ Explicitly, for
either integer or semi-integer $j$,%
\begin{equation}
\chi_{j}\left(  \theta\right)  =\frac{\sin\left(  \left(  2j+1\right)
\theta/2\right)  }{\sin\left(  \theta/2\right)  }\ .
\end{equation}
These characters are all real. \ Therefore the number of times that spin $s$
appears in the product $j_{1}\otimes\cdots\otimes j_{n}$ is%
\begin{equation}
M\left(  s,j_{1},\cdots,j_{n}\right)  =\frac{1}{\pi}\int_{0}^{2\pi}\chi
_{s}\left(  2\vartheta\right)  \chi_{j_{1}}\left(  2\vartheta\right)
\cdots\chi_{j_{n}}\left(  2\vartheta\right)  \sin^{2}\vartheta~d\vartheta\ ,
\label{IntegralForm}%
\end{equation}
where we have taken $\theta=2\vartheta$ to avoid having half-angles appear in
the invariant measure and the Chebyshev polynomials (e.g., see \cite{V}
Chapter III \S 8.1) thereby mapping the $SU\left(  2\right)  $ group manifold
$0\leq\theta\leq4\pi$ to $0\leq\vartheta\leq2\pi$. \ To re-emphasize earlier
remarks, we note that (\ref{IntegralForm}) is totally symmetric in $\left\{
s,j_{1},\cdots,j_{n}\right\}  $ and valid if $s$ or any of the $j$s are
integer or semi-integer, and we also note that $M\left(  s,j_{1},\cdots
,j_{n}\right)  =M\left(  0,s,j_{1},\cdots,j_{n}\right)  $. \ In general
$M\left(  s,j_{1},\cdots,j_{n}\right)  $ will obviously reduce to a finite sum
of integers through use of the Chebyshev
\href{https://en.wikipedia.org/wiki/Chebyshev_polynomials#Products_of_Chebyshev_polynomials}{product
identity}, $U_{m}U_{n}=\sum_{k=0}^{n}U_{m-n+2k}$ for $m\geq n$.
\ Alternatively, the integral form (\ref{IntegralForm}) for the multiplicity
always reduces to a finite sum of hypergeometric functions (e.g. see
(\ref{Hyper1}) and (\ref{Hyper2})\ to follow).

In particular, for $j_{1}=\cdots=j_{n}=j$, the $n$-fold product $j^{\otimes
n}$ can yield spin $s$ a number of times, as given by%
\begin{equation}
M\left(  s;n;j\right)  =\frac{1}{\pi}\int_{0}^{2\pi}\sin\left(  \left(
2s+1\right)  \vartheta\right)  \left(  \frac{\sin\left(  \left(  2j+1\right)
\vartheta\right)  }{\sin\left(  \vartheta\right)  }\right)  ^{n}\sin
\vartheta~d\vartheta\ , \label{ProdMultInt}%
\end{equation}
for $s,j\in\left\{  0,\frac{1}{2},1,\frac{3}{2},2,\cdots\right\}  $. \ Yet
again, we note that $M\left(  j;n;j\right)  =M\left(  0;n+1;j\right)  $.
\ Moreover, the symmetry of the integrand in (\ref{ProdMultInt}) permits us to
write%
\begin{equation}
M\left(  s;n;j\right)  =\int_{0}^{2\pi}\frac{\exp\left(  2is\vartheta\right)
}{2\pi}\left(  \frac{\sin\left(  \left(  2j+1\right)  \vartheta\right)  }%
{\sin\left(  \vartheta\right)  }\right)  ^{n}~d\vartheta-\int_{0}^{2\pi}%
\frac{\exp\left(  2i\left(  s+1\right)  \vartheta\right)  }{2\pi}\left(
\frac{\sin\left(  \left(  2j+1\right)  \vartheta\right)  }{\sin\left(
\vartheta\right)  }\right)  ^{n}~d\vartheta\ .
\end{equation}
Each integral in the last expression reduces to a simple residue,%
\begin{equation}
\int_{0}^{2\pi}\frac{\exp\left(  2is\vartheta\right)  }{2\pi}\left(
\frac{\sin\left(  \left(  2j+1\right)  \vartheta\right)  }{\sin\left(
\vartheta\right)  }\right)  ^{n}~d\vartheta=\frac{1}{2\pi i}\oint
z^{2s}\left(  \frac{z^{2j+1}-z^{-2j-1}}{z-z^{-1}}\right)  ^{n}\frac{dz}%
{z}=c_{0}\left(  s,n,j\right)  \ ,
\end{equation}
where $c_{k}$ are the coefficients in the Laurent expansion of the integrand,%
\begin{equation}
z^{2s}\left(  \frac{z^{2j+1}-z^{-2j-1}}{z-z^{-1}}\right)  ^{n}=z^{2s}\left(
\sum_{m=0}^{2j}z^{2\left(  m-j\right)  }\right)  ^{n}=\sum_{k=-2\left(
jn-s\right)  }^{2\left(  jn+s\right)  }z^{k}~c_{k}\left(  s,n,j\right)  \ .
\end{equation}
That is to say, $c_{0}$ is the coefficient of $z^{-2s}$ (or of $z^{+2s}$) in
the Laurent expansion of $(z^{-2j}+z^{-2j+2}+\cdots\allowbreak+z^{2j-2}%
+z^{2jn})$ \cite{Katriel}, a coefficient that is easily obtained, e.g. using
either Maple$^{\textregistered}$ or Mathematica$^{\textregistered}$. \ 

\subsection*{Explicit $SU\left(  2\right)  $ Results as Binomial Coefficients}

So then, the multiplicity is always given by a difference,
\begin{equation}
M\left(  s;n;j\right)  =c_{0}\left(  s,n,j\right)  -c_{0}\left(
s+1,n,j\right)  \ , \label{MIsADifference}%
\end{equation}
where $2s$ is any integer such that $0\leq2s\leq2nj$, and where $s=0$ is
always allowed when $j$ is an integer but is only allowed for even $n$ when
$j$ is a semi-integer. \ To be more explicit, the expansion of $\left(
z^{-2j}+z^{-2j+2}+\cdots+z^{2j-2}+z^{2j}\right)  ^{n}$ involves so-called
\textquotedblleft generalized binomial coefficients\textquotedblright\ (see
Eqn(3) in \cite{Bollinger}) which can be written as sums of products of the
usual binomial coefficients. \ Eventually (see Lemma 6 in \cite{Kirillov} and
the Appendix in \cite{Mendonca}) this leads to%
\begin{equation}
c_{0}\left(  s,n,j\right)  =\sum_{k=0}^{\left\lfloor \frac{nj+s}%
{2j+1}\right\rfloor }\left(  -1\right)  ^{k}\binom{n}{k}\binom{nj+s-\left(
2j+1\right)  k+n-1}{nj+s-\left(  2j+1\right)  k}\ .
\end{equation}
For example, if $j=1/2$ the $c_{0}$s reduce to a single binomial coefficient
\cite{Bethe}.%
\begin{equation}
c_{0}\left(  s,n,1/2\right)  =\binom{n}{n/2-s}\ ,\ \ \ M\left(
s;n;1/2\right)  =\binom{n}{n/2-s}-\binom{n}{n/2-s-1}\ , \label{SpinHalf}%
\end{equation}
where $0\leq2s\leq n$, with $s=0$ allowed only for even $n$.

\subsection*{A Lattice of Multiplicities}

One may visualize $M\left(  s;n;j\right)  $ as a 3-dimensional semi-infinite
lattice of points $\left(  s;n;j\right)  $ with integer multiplicities
appropriately assigned to each lattice point. \ There are many straight lines
on this lattice such that the multiplicities are polynomial in the line
parameterization. \ For example, along some of the lattice diagonals,%
\begin{equation}
M\left(  n;n;1\right)  =1\ ,\ \ \ M\left(  n-1;n;1\right)
=n-1\ ,\ \ \ M\left(  n-2;n;1\right)  =\tfrac{1}{2}~n\left(  n-1\right)  \ .
\end{equation}
These are, respectively, the number of ways the highest possible spin (i.e.
$s=n$), the 2nd highest spin ($s=n-1$), and the 3rd highest spin ($s=n-2$)
occur in the Kronecker product of $n$ vector (i.e. $s=1$) representations.
\ The form for the number of spins farther below the maximum $s=n$, that occur
in products of $n$ vectors, is%
\begin{subequations}
\begin{align}
M\left(  n-\left(  2k+2\right)  ;n;1\right)   &  =\tfrac{1}{\left(
2k+2\right)  !}~n\left(  n-1\right)  \left(  n-2\right)  \cdots\left(
n-k\right)  \times p_{k+1}\left(  n\right)  \ ,\\
M\left(  n-\left(  2k+3\right)  ;n;1\right)   &  =\tfrac{1}{\left(
2k+3\right)  !}~n\left(  n-1\right)  \left(  n-2\right)  \cdots\left(
n-k\right)  \times q_{k+2}\left(  n\right)  \ ,
\end{align}
for $k=0,1,2,3,\cdots$, where $p_{k+1}$ and $q_{k+2}$ are polynomials in $n$
of order $k+1$ and $k+2$, as follows.
\end{subequations}
\begin{subequations}
\begin{align}
p_{k+1}\left(  n\right)   &  =n^{k+1}+\tfrac{1}{2}\left(  k+1\right)  \left(
5k-2\right)  n^{k}+\tfrac{1}{24}\left(  k\right)  \left(  k+1\right)  \left(
75k^{2}-205k-134\right)  n^{k-1}+\cdots\ ,\\
q_{k+2}\left(  n\right)   &  =n^{k+2}+\tfrac{1}{2}\left(  k\right)  \left(
5k+7\right)  n^{k+1}+\tfrac{1}{24}\left(  k+1\right)  \left(  75k^{3}%
-85k^{2}-410k-168\right)  n^{k}+\cdots\ .
\end{align}
As an exercise, the reader may verify the complete polynomials for orders $2$,
$3$, $4$, and $5$.%
\end{subequations}
\begin{gather}
p_{2}\left(  n\right)  =n^{2}+3n-22\ ,\ \ \ q_{2}\left(  n\right)
=n^{2}-7\ ,\\
p_{3}\left(  n\right)  =n^{3}+12n^{2}-61n-192\ ,\ \ \ q_{3}\left(  n\right)
=n^{3}+6n^{2}-49n+6\ ,\nonumber\\
p_{4}\left(  n\right)  =n^{4}+26n^{3}-37n^{2}-1622n+120\ ,\ \ \ q_{4}\left(
n\right)  =n^{4}+17n^{3}-91n^{2}-587n+1200\ ,\nonumber\\
p_{5}\left(  n\right)  =n^{5}+45n^{4}+205n^{3}-5565n^{2}%
-17486n+48720\ ,\ \ \ q_{5}\left(  n\right)  =n^{5}+33n^{4}-23n^{3}%
-3393n^{2}+2542n+21000\ .\nonumber
\end{gather}
At the time of writing, the authors have not managed to identify the $p_{k}$
and $q_{k}$ polynomial sequences with any that were previously studied.

\subsection*{Tabulating Some Examples}

For more explicit examples, we tabulate the number of singlets that appear in
products $j^{\otimes n}$ for $j=1,\cdots,9$ and for $n=1,\cdots,10$. \ The
Table entries below were obtained just by evaluation of the integrals in
(\ref{ProdMultInt}) for $s=0$.%
\[%
\begin{array}
[c]{cccccccccc}%
\mathsf{M}\left(  \mathsf{0;n;j}\right)  &
\text{\text{\href{https://oeis.org/A005043}{\text{j = 1}}}} &
\text{\text{\href{https://oeis.org/A007043}{\text{j = 2}}}} &
\text{\text{\href{https://oeis.org/A264608}{\text{j = 3}}}} &
\text{\text{\href{https://oeis.org/A272393}{\text{j = 4}}}} &
\text{\text{\href{https://oeis.org/A272395}{\text{j = 5}}}} & \mathsf{j=6} &
\mathsf{j=7} & \mathsf{j=8} & \mathsf{j=9}\\
\mathsf{n=1} & 0 & 0 & 0 & 0 & 0 & 0 & 0 & 0 & 0\\
\mathsf{n=2} & 1 & 1 & 1 & 1 & 1 & 1 & 1 & 1 & 1\\
\mathsf{n=3} & 1 & 1 & 1 & 1 & 1 & 1 & 1 & 1 & 1\\
\mathsf{n=4} & 3 & 5 & 7 & 9 & 11 & 13 & 15 & 17 & 19\\
\text{\textsf{\text{\href{https://oeis.org/A005891}{\text{n = 5}}}}} & 6 &
16 & 31 & 51 & 76 & 106 & 141 & 181 & 226\\
\text{\textsf{\text{\href{https://oeis.org/A005917}{\text{n = 6}}}}} & 15 &
65 & 175 & 369 & 671 & 1105 & 1695 & 2465 & 3439\\
\mathsf{n=7} & 36 & 260 & 981 & 2661 & 5916 & 11\,516 & 20\,385 & 33\,601 &
52\,396\\
\mathsf{n=8} & 91 & 1085 & 5719 & 19\,929 & 54\,131 & 124\,501 & 254\,255 &
474\,929 & 827\,659\\
\mathsf{n=9} & 232 & 4600 & 33\,922 & 151\,936 & 504\,316 & 1370\,692 &
3229\,675 & 6836\,887 & 13\,315\,996\\
\mathsf{n=10} & 603 & 19\,845 & 204\,687 & 1178\,289 & 4779\,291 &
15\,349\,893 & 41\,729\,535 & 100\,110\,977 & 217\,915\,579
\end{array}
\]

\subsubsection*{Nonpolynomial Columns}

The columns of the Table are \emph{not} expressible as polynomials in $n$, for
any fixed $j$, but they may be written as sums of hypergeometric or rational
functions of $n$. \ For example, the first two columns may be written as%
\begin{align}
M\left(  0;n;1\right)   &  =3^{n}\sum_{k=0}^{n}\binom{n}{k}\binom{2k+1}%
{k+1}\left(  -\frac{1}{3}\right)  ^{k}=\frac{4^{n}}{\sqrt{\pi}}\frac
{\Gamma\left(  \frac{1}{2}+n\right)  }{\Gamma\left(  2+n\right)  }\left.
_{2}F_{1}\right.  \left(  -n,-1-n;\frac{1}{2}-n;\frac{1}{4}\right)
\ ,\label{Hyper1}\\
M\left(  0;n;2\right)   &  =\frac{1}{2}\sum_{k=0}^{n}\frac{\left(  -6\right)
^{k}~\Gamma\left(  \frac{1}{2}+k\right)  }{\Gamma\left(  1+\frac{k}{2}\right)
\Gamma\left(  \frac{3}{2}+\frac{k}{2}\right)  }\binom{n}{k}\left.  _{3}%
F_{2}\right.  \left(  \frac{1}{4}+\frac{k}{2},\frac{3}{4}+\frac{k}%
{2},k-n;1+\frac{k}{2},\frac{3}{2}+\frac{k}{2};-16\right)  \ . \label{Hyper2}%
\end{align}
To obtain these and other multiplicities as hypergeometric functions, for
integer $s$ and $j$, it is useful to change variables, to $t=\cos^{2}%
\vartheta$, so that (\ref{ProdMultInt}) becomes
\begin{equation}
M\left(  s;n;j\right)  =\frac{2}{\pi}~4^{s+nj}\int_{0}^{1}\left(
\prod\limits_{k=1}^{s}\left(  t-r_{k}\left(  s\right)  \right)  \right)
\left(  \prod\limits_{l=1}^{j}\left(  t-r_{l}\left(  j\right)  \right)
\right)  ^{n}\sqrt{\frac{1-t}{t}}~dt\ . \label{HyperForm}%
\end{equation}
The products here involve the known roots $r_{l}\left(  j\right)  $ of the
Chebyshev polynomials. \ For integer $j$,
\begin{equation}
U_{2j}\left(  \cos\left(  \vartheta\right)  \right)  =4^{j}\prod
\limits_{l=1}^{j}\left(  t-r_{l}\left(  j\right)  \right)  \ ,\ \ \ t=\cos
^{2}\vartheta\ ,\ \ \ r_{l}\left(  j\right)  =\cos^{2}\left(  \frac{l\pi
}{2j+1}\right)  \ , \label{ChebPolys&Roots}%
\end{equation}
while for semi-integer $j$, for comparison to the integer case,
\begin{equation}
U_{2j}\left(  \cos\left(  \vartheta\right)  \right)  =4^{j}\sqrt{t}%
~\prod\limits_{l=1}^{j-\frac{1}{2}}\left(  t-\ r_{l}\left(  j\right)  \right)
\ ,
\end{equation}
with the usual convention that the empty product is $1$.

The columns of the Table should be compared to the multiplicities of integer
spins that appear in the product of $2m$ spin $1/2$ representations. \ These
are well-known to be given by the
\href{https://en.wikipedia.org/wiki/Catalan's_triangle}{Catalan triangle}
\cite{SU(2)q},%
\begin{equation}
M\left(  s;2m;1/2\right)  =\frac{\left(  1+2s\right)  \left(  2m\right)
!}{\left(  m-s\right)  !\left(  m+s+1\right)  !}\ , \label{Catalan}%
\end{equation}
as follows from (\ref{SpinHalf}). \ As an aside, it is perhaps not so
well-known that multiplicities of all $SU\left(  N\right)  $ representations
occurring in the product of $n$ fundamental $N$-dimensional representations
are given by \href{http://oeis.org/A005789}{$N$-dimensional Catalan
structures} \cite{MultiCat,SU(N)}. \ 

Be that as it may, this aside suggests an alternate route to obtain and to
re-express some of the above results, especially for $j=1$, a route that
\emph{retraces} [pun intended] many of the logical steps. \ This other route
uses the explicit formula \cite{SU(N)} for products of fundamental triplets of
the group $SU\left(  3\right)  $\ and the \textquotedblleft tensor
embedding\textquotedblright\ $SU\left(  3\right)  \supset SU\left(  2\right)
$ (where the triplet of $SU\left(  3\right)  $ is identified with the $s=1$
vector representation) to deduce the number of $s=0$ singlets appearing in the
product of $n$ vector representations of $SU\left(  2\right)  $, namely,
\begin{equation}
M\left(  0;n;1\right)  =\left(  -1\right)  ^{n}\left.  _{2}F_{1}\right.
\left(  -n,\tfrac{1}{2};2;4\right)  \ .
\end{equation}
This is in exact agreement with the seemingly different result (\ref{Hyper1}).
\ Combining this with the elementary recursion relation that follows from
$\overrightarrow{s}\otimes\overrightarrow{1}=\overrightarrow{s+1}%
\oplus\overrightarrow{s}\oplus\overrightarrow{s-1}$, namely,%
\begin{equation}
M\left(  s;n;1\right)  =M\left(  s+1;n-1;1\right)  +M\left(  s;n-1;1\right)
+M\left(  s-1;n-1;1\right)  \ ,
\end{equation}
one then obtains $M\left(  s;n;1\right)  $ as a sum of Gauss hypergeometric
functions. \ Relations between contiguous functions then simplify the result
to a single hypergeometric function,%
\begin{equation}
M\left(  s;n;1\right)  =\left(  -1\right)  ^{n+s}\binom{n}{s}\left.  _{2}%
F_{1}\right.  \left(  s-n,s+\frac{1}{2};2+2s;4\right)  \ .
\end{equation}
Finally, the standard integral representation for $\left.  _{2}F_{1}\right.  $
eventually leads to the same integral form for $M\left(  s;n;1\right)  $ as
given by (\ref{ProdMultInt}) for $j=1$.

\subsubsection*{Polynomial Rows}

In contrast to the columns, the rows of the Table \emph{are} expressible as
polynomials in $j$ for any fixed $n$. \ Starting with $n=3$, the entries in
the $n$th row of the Table are polynomials in $j$ of order $n-3$. \ The fourth
row is obviously just the dimension of the spin $j$ representation, and the
fifth row is less obviously $1+\frac{5}{2}c_{j}$, where $c_{j}$ is the
quadratic $su\left(  2\right)  $ Casimir for spin $j$. \ In fact, based on the
numbers displayed above and some modest extensions of the Table, the row
entries are seen to be of the form $poly_{\left(  n-3\right)  /2}\left(
c_{j}\right)  $ for odd $n\geq3$ and $poly_{\left(  n-4\right)  /2}\left(
c_{j}\right)  \times d_{j}$ for even $n\geq4$, where $poly_{k}\left(
c\right)  $ is a polynomial in $c$ of order $k$. \ For the\ last eight rows of
the Table these polynomials are given by:%
\begin{gather}%
\begin{array}
[c]{ccccc}%
\mathsf{n=3} & 1\medskip &  & \mathsf{n=4} & d_{j}\medskip\\
\mathsf{n=5} & 1+\frac{5}{2}c_{j}\medskip &  & \mathsf{n=6} & \left(
1+2c_{j}\right)  d_{j}\medskip\\
\mathsf{n=7} & 1+\frac{14}{3}c_{j}+\frac{77}{12}c_{j}^{2}\medskip &  &
\mathsf{n=8} & \left(  1+4c_{j}+\frac{16}{3}c_{j}^{2}\right)  d_{j}\medskip\\
\mathsf{n=9} & 1+\frac{27}{4}c_{j}+\frac{73}{4}c_{j}^{2}+\frac{289}{16}%
c_{j}^{3}\medskip &  & \mathsf{n=10} & \left(  1+6c_{j}+\frac{143}{9}c_{j}%
^{2}+\frac{140}{9}c_{j}^{3}\right)  d_{j}\medskip
\end{array}
\\
\text{where\ }d_{j}=1+2j\ ,\ \ \ \text{and\ \ \ }c_{j}=j\left(  1+j\right)
\ .
\end{gather}
Thus the\ ten rows of the Table may be effortlessly extended to arbitrarily
large $j$. \ Moreover, to obtain the polynomial that gives any row for $n>10$,
for arbitrary values of $j$, it is only necessary to evaluate $M\left(
0;n;j\right)  $ for $1\leq j\leq\left\lfloor \frac{n-1}{2}\right\rfloor $.
\ Once again, at the time of writing, the authors have not managed to identify
this polynomial sequence with any that were previously studied.

\subsection*{Asymptotic Behavior}

Finally, consider the extension of the columns of the Table to arbitrarily
large $n$, or more generally, consider the asymptotic behavior of $M\left(
s;n;j\right)  $ as $n\rightarrow\infty$ for fixed $s$ and $j$. \ This behavior
can be determined in a straightforward way, for any $s$ and $j$, by a careful
asymptotic analysis of the integral in (\ref{ProdMultInt}). \ Such
$n\rightarrow\infty$ behavior may be of interest in various statistical problems.

The simplest illustration is $M\left(  0;n;1/2\right)  $ for even $n$. \ For
this particular case, (\ref{Catalan}) and Stirling's approximation,
$n!\underset{n\rightarrow\infty}{\sim}\sqrt{2\pi n}\left(  \frac{n}{e}\right)
^{n}$, give directly the main term in the asymptotic behavior,%
\begin{equation}
M\left(  0;2m;1/2\right)  \underset{m\rightarrow\infty}{\sim}\frac{4^{m}%
}{m^{3/2}\sqrt{\pi}}\left(  1+O\left(  \frac{1}{m}\right)  \right)  \ .
\label{SpinHalfSingAsymp}%
\end{equation}
On the other hand, upon setting $t=\cos^{2}\vartheta$ the integral
(\ref{ProdMultInt}) has a form like that in (\ref{HyperForm}), namely,%
\begin{equation}
M\left(  0;2m;1/2\right)  =\frac{2}{\pi}~4^{m}\int_{0}^{1}t^{m}\sqrt
{\frac{1-t}{t}}~dt=\frac{2}{\pi}~4^{m}B\left(  m+\frac{1}{2},\frac{3}%
{2}\right)  \ . \label{SpinHalfSingInt}%
\end{equation}
The $t$ integral is just a beta function, $B\left(  m+\frac{1}{2},\frac{3}%
{2}\right)  =\Gamma\left(  m+\frac{1}{2}\right)  \Gamma\left(  \frac{3}%
{2}\right)  /\Gamma\left(  m+2\right)  $, which leads back to exactly
(\ref{Catalan}) for $s=0$. \ But rather than using Stirling's approximation,
it is more instructive to determine the asymptotic behavior directly from the
integral (\ref{SpinHalfSingInt}) using
\href{https://en.wikipedia.org/wiki/Watson's_lemma}{Watson's lemma}. \ Thus%
\begin{equation}
M\left(  0;2m;1/2\right)  \underset{m\rightarrow\infty}{\sim}2\sqrt{2}%
~\frac{2^{2m}}{\left(  2m\right)  ^{3/2}\sqrt{\pi}}\left(  1-\frac{9}%
{8m}+O\left(  \frac{1}{m^{2}}\right)  \right)  \ .
\end{equation}
Naively it might be expected that\emph{ }the leading asymptotic behavior
(\ref{SpinHalfSingAsymp}) follows from a heuristic, saddle-point-Gaussian
evaluation of the integration in (\ref{SpinHalfSingInt}). \ Unfortunately,
that expectation is not fulfilled.\ \ The correct $m$ dependence is obtained
for $M$, but with an incorrect overall coefficient. \ To obtain the correct
coefficient, a more careful analysis of the asymptotic behavior is needed, as
provided by Watson's lemma.

Similarly, for large $n$ the number of singlets occurring in the product of
$n$ spin $1$ representations behaves as%
\begin{equation}
M\left(  0;n;1\right)  \underset{n\rightarrow\infty}{\sim}\frac{3\sqrt{3}}%
{8}~\frac{3^{n}}{n^{3/2}\sqrt{\pi}}\left(  1-\frac{21}{16n}+O\left(  \frac
{1}{n^{2}}\right)  \right)  \ , \label{SpinOneSingAsymp}%
\end{equation}
and the number of singlets in the product of $n$ spin $2$ representations
behaves as%
\begin{equation}
M\left(  0;n;2\right)  \underset{n\rightarrow\infty}{\sim}\frac{1}{8}%
~\frac{5^{n}}{n^{3/2}\sqrt{\pi}}\left(  1-\frac{15}{16n}+O\left(  \frac
{1}{n^{2}}\right)  \right)  \ . \label{SpinTwoSingAsymp}%
\end{equation}

In general, the number of spin $s$ representations occurring in the product of
$n$ spin $j$ representations for large $n$ has asymptotic behavior \cite{ADF}
\begin{equation}
M\left(  s;n;j\right)  \underset{n\rightarrow\infty}{\sim}\left(  1+2s\right)
\left(  \frac{3}{2j\left(  j+1\right)  }\right)  ^{3/2}~\frac{\left(
1+2j\right)  ^{n}}{n^{3/2}\sqrt{\pi}}\left(  1-\frac{3}{4n}-\frac{9}{8n}%
\frac{1}{j\left(  j+1\right)  }-\frac{3}{2n}\frac{s\left(  s+1\right)
}{j\left(  j+1\right)  }+O\left(  \frac{1}{n^{2}}\right)  \right)  \ .
\label{SpinJSpinSAsymp}%
\end{equation}
This is correct for either integer or semi-integer $s$ or $j$, although of
course $n$ must be (odd) even to obtain (semi-)integer $s$ from products of
semi-integer $j$, and only integer $s$ are produced by integer $j$.
\ Asymptotically then, for integer $j$,
\begin{equation}
M\left(  j;n;j\right)  /M\left(  0;n;j\right)  =M\left(  0;n+1;j\right)
/M\left(  0;n;j\right)  \underset{n\rightarrow\infty}{\sim}1+2j+O\left(
\frac{1}{n}\right)  \ .
\end{equation}
Remarkably, this behavior is approximately seen in the Table, with errors
$\lessapprox10\%$. \ On the other hand, for semi-integer $j$ and even $n$,%
\begin{equation}
M\left(  j;n+1;j\right)  /M\left(  0;n;j\right)  =M\left(  0;n+2;j\right)
/M\left(  0;n;j\right)  \underset{n\rightarrow\infty}{\sim}\left(
1+2j\right)  ^{2}+O\left(  \frac{1}{n}\right)  \ .
\end{equation}

For integer $j$, the result in (\ref{SpinJSpinSAsymp}) follows directly,
albeit tediously, from an application of Watson's lemma to (\ref{HyperForm})
after switching to exponential variables. \ In that case the overall
coefficient in (\ref{SpinJSpinSAsymp}) arises as a simple algebraic function
of the Chebyshev roots in (\ref{ChebPolys&Roots}), namely, $1/\left(
\sum_{l=1}^{j}\frac{1}{1-r_{l}\left(  j\right)  }\right)  ^{3/2}$. \ This then
reduces to the Casimir-dependent expression in (\ref{SpinJSpinSAsymp}) by
virtue of the integer $j$ identity,%
\begin{equation}
\sum_{l=1}^{j}\frac{1}{1-r_{l}\left(  j\right)  }=\frac{2}{3}~j\left(
j+1\right)  \ . \label{C(j)}%
\end{equation}
Similar statements apply when $j$ is semi-integer leading again to
(\ref{SpinJSpinSAsymp}). \ For semi-integer $j$ the relevant identity is%
\begin{equation}
\frac{1}{2}+\sum_{l=1}^{j-1/2}\frac{1}{1-r_{l}\left(  j\right)  }=\frac{2}%
{3}~j\left(  j+1\right)  \ ,
\end{equation}
with the usual convention that the empty sum is $0$.

\subsection*{All-Order Extensions of the Asymptotics}

The asymptotic behavior given by (\ref{SpinJSpinSAsymp}) is useful for fixed
$s$ and $j$ in the limit as $n\rightarrow\infty$. \ If the resulting spin $s$
produced by the $n$-fold product is also allowed to become large in the limit,
e.g. $s=O\left(  \sqrt{n}\right)  $, then (\ref{SpinJSpinSAsymp})\ is
\emph{not} useful. \ However, in that particular case it is possible to use
renomalization group methods \cite{RG} to sum the series of terms involving
powers of $\frac{1}{n}\frac{s\left(  s+1\right)  }{j\left(  j+1\right)  }$ to
obtain an exponential, and hence an improved approximation. \ The result is%
\begin{equation}
M\left(  s;n;j\right)  \underset{n\rightarrow\infty}{\sim}\left(  1+2s\right)
\left(  \frac{3}{2j\left(  j+1\right)  }\right)  ^{3/2}~\frac{\left(
1+2j\right)  ^{n}}{n^{3/2}\sqrt{\pi}}~e^{-\frac{3}{2n}\frac{s\left(
s+1\right)  }{j\left(  j+1\right)  }}~\left(  1-\frac{3}{4n}-\frac{9}{8n}%
\frac{1}{j\left(  j+1\right)  }+O\left(  \frac{1}{n^{2}}\right)  \right)  \ .
\label{TSvKAsymptotics}%
\end{equation}
For large $n$ this last expression gives
\href{https://cgc.physics.miami.edu/SpinAsymptotics.html}{an excellent
approximation} out to values\ of $s$ of order $\sqrt{n}$ and beyond.
\ Moreover, the peak in the distribution of spins $s$ produced by the product
of $n$ spin $j$s is given for large $n$ by
\begin{equation}
s_{\text{mult}}\underset{n\rightarrow\infty}{\sim}\sqrt{nj\left(  j+1\right)
/3}\ . \label{peak}%
\end{equation}
This follows from the exact result (\ref{Catalan}) for spin $1/2$, or from
(\ref{TSvKAsymptotics}) for any $j$. \ Alternatively, for specific $j$ the
direct numerical evaluation of either (\ref{ProdMultInt}) or
(\ref{MIsADifference}) verifies (\ref{peak}) upon taking $n$ large, say,
$n\approx10^{4}$.

Perhaps some further insight is provided by the asymptotic behavior of the
continuous function that gives the \emph{normalized number of states with a
given total spin}, $s$, as obtained from (\ref{TSvKAsymptotics}). \ This is \
\begin{equation}
\frac{\left(  1+2s\right)  M\left(  s;n;j\right)  }{\left(  1+2j\right)  ^{n}%
}~dj\underset{n\rightarrow\infty}{\sim}\left(  1-\frac{3}{4n}\left(
1+\frac{1}{s\left(  s+1\right)  }\right)  +O\left(  \frac{1}{n^{2}}\right)
\right)  ~P\left(  x\right)  ~dx\ , \label{ChiSquared}%
\end{equation}
where, with a suitable choice of the variable $x$, $P\left(  x\right)  $ is
the normalized
\href{https://en.wikipedia.org/wiki/Chi-squared_distribution}{chi-squared
probability distribution function} for \emph{three} degrees of freedom:%
\begin{equation}
x\equiv\frac{3\left(  1+2j\right)  ^{2}}{8ns\left(  s+1\right)  }%
\ ,\ \ \ P\left(  x\right)  =\frac{2}{\sqrt{\pi}}~\sqrt{x}~e^{-x}%
\ ,\ \ \ \int_{0}^{\infty}P\left(  x\right)  dx=1\ .
\end{equation}
In retrospect, this may not be a total surprise since the underlying rotation
group may be parameterized by \emph{three} Euler angles. \ Note that this last
asymptotic form is correctly normalized to give the total number of states as
$n\rightarrow\infty$, i.e.
\begin{equation}
\lim_{n\rightarrow\infty}\frac{1}{\sqrt{\pi}}\int_{0}^{nj}\left(  1+2s\right)
^{2}~\left(  \frac{3}{2nj\left(  j+1\right)  }\right)  ^{3/2}e^{-\frac{3}%
{2n}\frac{s\left(  s+1\right)  }{j\left(  j+1\right)  }}~ds=1\ .
\end{equation}
Also note that the expression for the number of states, (\ref{ChiSquared}),
has a maximum at spin%
\begin{equation}
s_{\text{state}}\underset{n\rightarrow\infty}{\sim}\sqrt{2}~s_{\text{mult}%
}\underset{n\rightarrow\infty}{\sim}\sqrt{2nj\left(  j+1\right)  /3}\text{ .}%
\end{equation}
\ 

\section*{Some Applications}

In closing, we stress that spin multiplicities play useful roles in a wide
range of fields, too numerous to present in detail here. \ But we briefly
sketch a few applications of the results described above.

Some of the $SU\left(  2\right)  $ results for $s=0$ have been used for
decades in elasticity theory \cite{Ogden} and in quantum chemistry \cite{AT},
as well as in nuclear physics, as is evident from the literature we have cited
upon recognizing that
\href{http://mathworld.wolfram.com/IsotropicTensor.html}{the number of
isotropic rank-$n$ tensors} in three dimensions is just $M\left(
0;n;1\right)  $.

The theory of group characters has been widely used in lattice gauge theory
calculations for a long time \cite{BGZ,Creutz} and continues to play an
important role in various strong coupling calculations \cite{Unger}.
\ Characters are also indispensible to determine the spin content of various
string theories \cite{CGGT}$.$

More generally, generic representation composition results continually find
new uses. \ Recent examples include frustration and entanglement entropy for
spin chains, with possible applications to black hole physics \cite{Shor}.

Multiplicities such as those in the Table have also attracted some recent
attention in the field of quantum computing, ultimately with implications for
cryptography. \ In particular, there are so-called \textquotedblleft
entanglement witness\textquotedblright\ (EW) operators that allow the
detection of entangled states \cite{LKCH,Toth,CHI}. \ By knowing the
degeneracy of the EW eigenstates for an $n$-particle state, one can determine
the fraction of all states for which entanglement is \textquotedblleft
decidable\textquotedblright\ --- a fraction that is especially of interest in
the limit of large $n$. For systems of $n$ spin $j$ particles, with the EW
operator taken to be the Casimir of the total spin, this fraction of decidable
states \cite{CHI} is denoted $f_{j}\left(  n\right)  $. \ In this case, from
the asymptotic expression given above in (\ref{ChiSquared}), one readily
obtains
\href{https://www.researchgate.net/publication/304660051_Decidable_States_in_the_Large_N_Limit}{the
exact result}%
\begin{equation}
\lim_{n\rightarrow\infty}~f_{j}\left(  n\right)  =f_{j}\left(  \infty\right)
=\operatorname{erf}\left(  \sqrt{\frac{3/2}{s+1}}\right)  -\sqrt{\frac{6/\pi
}{s+1}}~\exp\left(  -\frac{3/2}{s+1}\right)  \ ,
\end{equation}
where $\operatorname{erf}\left(  x\right)  =2\int_{0}^{x}\exp\left(
-s^{2}\right)  ds/\sqrt{\pi}$ is the conventional
\href{https://en.wikipedia.org/wiki/Error_function}{error function}. \ 

Many other statistical applications of spin multiplicities for large $n$ have
been proposed in a recent, independent investigation of this subject
\cite{Poly}.\bigskip

\textbf{Acknowledgements:} \ We thank J Katriel and J Mendon\c{c}a for
pointing out elegant ways to re-express the multiplicity in the general case.
\ We also thank A Polychronakos and K Sfetsos for an advance copy of their
paper. \ Finally, we thank an anonymous reviewer for bringing \cite{Kirillov}%
\ to our attention. \ This work was supported in part by a University of Miami
Cooper Fellowship.


\begin{thebibliography}{99}
\bibitem {Katriel}This was emphasized to us by J Katriel [private
communication]. \ Note that $c_{0}\left(  s,n,j\right)  $ is just the number
of states with $J_{z}=s$ (or with $J_{z}=-s$) that occur in $j^{\otimes n}$.
\bibitem {SU(N)}Explicit generalizations of (\ref{Catalan}) for $SU\left(
N\right)  $\ multiplicities are straightforward to obtain upon invoking the
equality of such multiplicities to the dimensions of representations of the
symmetric group, $S_{n}$, and then computing the latter as $n!/\left(
\prod~\text{hooks}\right)  $ after expressing the representations as Young diagrams.
\end{thebibliography}
\end{document}